\documentclass[runningheads]{llncs}
\usepackage[T1]{fontenc}
\usepackage{soul}
\usepackage{url}
\usepackage[hidelinks]{hyperref}
\usepackage[utf8]{inputenc}
\usepackage{tikz}
\usetikzlibrary{arrows.meta, positioning, fit, shapes}
\usepackage{graphicx}
\usepackage{color}

\usepackage{amsmath}
\usepackage{amssymb}
\usepackage{booktabs}
\usepackage{bm}
\usepackage{algorithm}
\usepackage{algorithmic}
\usepackage{bcprules}
\usepackage{chemarrow}
\usepackage{cite}

\usepackage{amsfonts}
\usepackage{xcolor}
\usepackage{ifthen}
\usepackage{nicefrac}
\usepackage{wrapfig}
\makeatletter
\let\MYcaption\@makecaption
\makeatother
\usepackage{subcaption}
\usepackage[normalem]{ulem}
\usepackage{wrapfig}
\usepackage{listings}
\usepackage[firstpage]{draftwatermark} 

\SetWatermarkAngle{0}
\SetWatermarkText{\raisebox{12.5cm}{
\hspace{0.1cm}
\href{https://doi.org/10.1109/5.771073}{\includegraphics{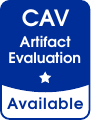}}
\hspace{9cm}
\includegraphics{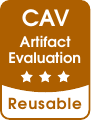}
}}

\allowdisplaybreaks[1]

\newcommand\Fig[1] {Fig.~\ref{#1}}

\newcommand\Sec[1] {Sect.~\ref{#1}}
\newcommand\Sects[1] {Sects.~\ref{#1}}

\newcommand\Eg[1] {Example~\ref{#1}}
\newcommand\Tbl[1] {Table~\ref{#1}}

\renewcommand{\phi} {\varphi}

\newcommand{\eqdef}{\ensuremath{\stackrel{\mathrm{def}}{=}}}
\newcommand{\detassign}{\ensuremath{:=}}

\newcommand{\mean}{\mathsf{mean}}

\newcommand{\Gospel}[0]{\textsf{Gospel}}
\newcommand{\WhyML}[0]{\text{WhyML}}

\newcommand{\StatWhy}[0]{\textsf{StatWhy}}

\newcommand{\WhyT}[0]{\textsf{Why3}}
\newcommand{\Cameleer}[0]{\textsf{Cameleer}}
\newcommand{\cvcFive}[0]{\textsc{cvc5}}

\newcommand{\ppl}[0]{}

\newcommand{\sideL}[0]{\mathsf{L}}
\newcommand{\sideU}[0]{\mathsf{U}}

\newcommand{\phiL}[0]{\phi_{\sideL}}
\newcommand{\phiU}[0]{\phi_{\sideU}}

\newcommand{\psiPre}[0]{\psi_{{\sf pre}}}

\newcommand{\phiPost}[0]{\phi_{{\sf post}}}

\newcommand{\Normal}[0]{\mathit{N}}

\newcommand{\alg}[0]{\mathit{A}}

\makeatletter
\providecommand{\leftsquigarrow}{%
  \mathrel{\mathpalette\reflect@squig\relax}%
}
\newcommand{\reflect@squig}[2]{%
  \reflectbox{$\m@th#1\rightsquigarrow$}%
}
\makeatother
\makeatletter
\newcommand{\subalign}[1]{%
  \vcenter{%
    \Let@ \restore@math@cr \default@tag
    \baselineskip\fontdimen10 \scriptfont\tw@
    \advance\baselineskip\fontdimen12 \scriptfont\tw@
    \lineskip\thr@@\fontdimen8 \scriptfont\thr@@
    \lineskiplimit\lineskip
    \ialign{\hfil$\m@th\scriptstyle##$&$\m@th\scriptstyle{}##$\hfil\crcr
      #1\crcr
    }%
  }%
}
\makeatother
\newcommand{\sampled}[3]{#1 \mathbin{\leftsquigarrow_{\substack{\\[0.1ex]\hspace{-1.5ex}#3\hspace{0.2ex}}\!}} #2}

\newcommand{\kappaX}[1]{\kappa_{#1}}

\newcommand{\cmpds}[1]{\varpi_{#1}}

\renewcommand{\implies}{\rightarrow}

\newcommand{\Know}{\mathbf{K}}
\newcommand{\Possible}{\mathbf{P}}

\newcommand{\KnowXx}[2]{\mathop{\mathbf{K}^{#1}_{#2}}}

\newcommand{\env}{\Gamma}

\newcommand{\triple}[3]{\{#1\}\ #2\ \{#3\}}

\newcommand{\axTT}{\textsc{Two-HT}}

\newcommand{\drugA}{\texttt{drug1}}
\newcommand{\drugB}{\texttt{drug2}}

\newif\ifcommentson\commentsonfalse

\newif\ifconferenceon\conferenceontrue
\ifconferenceon
\newcommand{\arxiv}[1]{}
\newcommand{\conference}[1]{#1}
\newcommand{\conferenceShort}[1]{}
\newcommand{\commentsize}[0]{.95\textwidth}
\else
\newcommand{\arxiv}[1]{#1}
\newcommand{\conference}[1]{}
\newcommand{\conferenceShort}[1]{}
\fi

\newif\ifanonymous\anonymousfalse

\ifcommentson
\newcommand{\commentYK}[1]{\begin{center} \parbox{\commentsize}{\textbf{\textcolor{black}{Comment Y.}} \textcolor{red}{#1} }\end{center}}
\newcommand{\commentKK}[1]{\begin{center} \parbox{\commentsize}{\textbf{\textcolor{black}{Comment KK.}} \textcolor{red}{#1 }}\end{center}}
\newcommand{\commentKS}[1]{\begin{center} \parbox{\commentsize}{\textbf{\textcolor{black}{Comment KS.}} \textcolor{red}{#1} }\end{center}}
\newcommand{\replyYK}[1]{\begin{center} \parbox{\commentsize}{\textbf{Reply Y.} \textcolor{blue}{#1} }\end{center}}
\newcommand{\replyKK}[1]{\begin{center} \parbox{\commentsize}{\textbf{Reply KK.} \textcolor{blue}{#1} }\end{center}}
\newcommand{\replyKS}[1]{\begin{center} \parbox{\commentsize}{\textbf{Reply KS.} \textcolor{blue}{#1} }\end{center}}
\newcommand{\commentY}[1]{\marginpar{\footnotesize \color{red} {\bf Y:} \textsf{\scriptsize #1}}}
\newcommand{\commentKk}[1]{\marginpar{\footnotesize \color{red} {\bf KK:} \textsf{\scriptsize #1}}}
\newcommand{\commentKs}[1]{\marginpar{\footnotesize \color{red} {\bf KS:} \textsf{\scriptsize #1}}}
\newcommand{\replyY}[1]{\marginpar{\footnotesize \color{red} {\bf Y:} \textsf{\scriptsize #1}}}
\newcommand{\replyKk}[1]{\marginpar{\footnotesize \color{red} {\bf KK:} \textsf{\scriptsize #1}}}
\newcommand{\replyKs}[1]{\marginpar{\footnotesize \color{red} {\bf KS:} \textsf{\scriptsize #1}}}
\else
\newcommand{\commentYK}[1]{}
\newcommand{\commentKK}[1]{}
\newcommand{\commentKS}[1]{}
\newcommand{\replyYK}[1]{}
\newcommand{\replyKK}[1]{}
\newcommand{\replyKS}[1]{}
\newcommand{\commentY}[1]{}
\newcommand{\commentKk}[1]{}
\newcommand{\commentKs}[1]{}
\newcommand{\replyY}[1]{}
\newcommand{\replyKk}[1]{}
\newcommand{\replyKs}[1]{}
\fi

\definecolor{DarkGreen}{rgb}{0,.6,0}

\newcommand{\colorR}[1]{\textcolor{red}{#1}}

\newcommand{\commentOut}[1]{}

\newcommand{\pagelimitmarker}[1]{~\\ {\colorR{\ifthenelse{\thepage>#1}{\Huge Exceeding the page limit}{\huge Within the page limit}}}~\\ {\huge{\colorR{~~Page Limit\,\,\,\,\, = #1}}}~\\ {\huge{\colorR{~~Current Page = $\thepage$}}}}

\begin{document}
\title{StatWhy: Formal Verification Tool for Statistical Hypothesis Testing Programs \thanks{The artifact of the paper is available at \texttt{\url{https://github.com/fm4stats/statwhy}} and \texttt{\url{https://zenodo.org/records/13991312}}.}}
\ifanonymous
\author{Anonymized}
\institute{Anonymized}
\else
\author{Yusuke Kawamoto\inst{1}
\orcidID{0000-0002-2151-9560}
\and
Kentaro Kobayashi\inst{1,2}
\and
Kohei Suenaga\inst{3}
\orcidID{0000-0002-7466-8789}
\thanks{The authors are listed in alphabetical order.}
}
\authorrunning{Y. Kawamoto et al.}
\institute{AIST, Tokyo, Japan \and
University of Tsukuba, Ibaraki, Japan \and
Kyoto University, Kyoto, Japan
}
\fi
\maketitle              
\begin{abstract}
Statistical methods have been widely misused and misinterpreted in various scientific fields, raising significant concerns about the integrity of scientific research.
To mitigate this problem,
we propose a 
tool-assisted
method for formally specifying and automatically verifying the correctness of statistical programs.
In this method, programmers are required to annotate the source code of the statistical programs with the requirements for these methods.
Through this annotation, they are reminded to check the requirements for statistical methods, including those that cannot be formally verified, such as the distribution of the unknown true population.
Our software tool \StatWhy{} automatically checks whether programmers have properly specified the requirements for the statistical methods, thereby identifying any missing requirements that need to be addressed.
This tool is implemented using the Why3 platform to verify the correctness of OCaml programs that conduct statistical hypothesis testing.
We demonstrate how \StatWhy{} can be used to avoid common errors in various statistical hypothesis testing programs.
\end{abstract}

\section{Introduction}
\label{sec:intro}

Statistical techniques have been essential for acquiring scientific knowledge from data
in various academic fields.
In particular, an increasing number of researchers have used \emph{statistical hypothesis testing}~\cite{Arbuthnot:1710,Hogg:2013:book-intro} to derive scientific conclusions from datasets. 
However, these statistical methods have been widely misused and misinterpreted,
raising significant concerns about the integrity of scientific research~\smash{\cite{Fernandes-Taylor:11:BMCRN,Makin:19:elife}}.
For example, the notion of \emph{statistical significance}, assessed by calculating \emph{$p$-values}, has been widely misused and misinterpreted~\cite{Wasserstein:16:AS}.

For this reason, various guidelines for statistical analyses \smash{\cite{von:07:STROBE,Moher:12:CONSORT}} have been proposed
to improve the quality and reproducibility of scientific research.
However, owing to the absence of a formal language to describe procedures,
we need to manually refer to these guidelines, written in natural language.  
As a result, the correctness of statistical analyses has not been checked automatically.

To mitigate these problems,
we propose a new method for the formal specification and automatic verification of statistical program correctness.
Specifically, programmers are required to annotate their source code with the requirements for the statistical methods and the interpretations of the analysis results. 
Then, our tool \StatWhy{} automatically checks whether these requirements and interpretations are correctly annotated.
For example, 
\StatWhy{} can verify whether a $p$-value is correctly calculated in a program, thus preventing $p$-value hacking, i.e., a technique to manipulate statistical analyses to obtain a lower $p$-value.

The goal of 
\StatWhy{}
stems from the nature of statistics:
many requirements for statistical methods cannot be proven mathematically
because they are usually properties of an unknown true population that analysts seek to estimate from sampled data.
For example, many statistical hypothesis testing methods require a population to follow a normal distribution.
Since analysts cannot prove this requirement mathematically,
they are responsible for judging whether the population appears to follow a normal distribution, possibly using their background knowledge about the population.
For this reason, \StatWhy{} asks analysts to explicitly write down the requirements for statistical methods%
---typically, the assumptions that they make about the population distributions---
as an annotation in their source code. 
Then, the analysts are reminded to check these requirements empirically and approximately using their background knowledge.

To design \StatWhy{}, we use the framework of belief Hoare logic (BHL)~\cite{Kawamoto:21:KR,Kawamoto:24:AIJ} 
and provide constructs to make writing statistical programs easier, as well as libraries for the specification of various hypothesis testing methods.
For the implementation of this tool, we rely on the \WhyT{} platform~\cite{Filliatre:13:ESOP}
to handle practical programming languages and to automatically discharge verification conditions using external SMT solvers.

Although the current implementation of \StatWhy{} focuses on statistical hypothesis testing,
the approach is not limited to a specific branch of statistics.  Rather, it can be applied to any situation where the usage of statistical methods in programs needs to be checked.
In future versions of the tool, we will include additional statistical methods beyond hypothesis testing.

\noindent
\textbf{Contributions.}
Our main contributions are summarized as follows:
\begin{enumerate}
\item We propose a new tool-assisted method for formally specifying and automatically verifying statistical programs 
(i.e., programs that perform hypothesis testing and calculate statistics under certain assumptions about an unknown population).
This method requires programmers to annotate their source code with the requirements and the interpretations of statistical analyses, 
which makes a statistical procedure verifiable and explainable.
\item We implemented a software tool \StatWhy{} based on our verification method. 
Given a program as input, \StatWhy{} automatically verifies whether a programmer has correctly annotated it with the requirements for statistical hypothesis testing and the interpretation of the test results.
\StatWhy{} is available with a range of examples and comprehensive documentation~\cite{Kawamoto:24:UD}.
\item We demonstrate how \StatWhy{} can be used to avoid common errors in various popular statistical hypothesis testing methods. 
\end{enumerate}
To the best of our knowledge, \StatWhy{} appears to be the first tool to automatically verify the requirements for the appropriate use of hypothesis tests.
This work represents the first step in building a framework for specifying and verifying the integrity of scientific conclusions based on statistical analyses.

\noindent
\textbf{Related Work.}
\label{sub:related}
\noindent
\textit{Logic for Statistics.}
Several studies on modal logic have been proposed to express statistical properties~\cite{Kawamoto:19:FC,Kawamoto:23:JELIA}.
The work on statistical epistemic logic ({StatEL})~\cite{Kawamoto:19:FC,Kawamoto:19:SEFM,Kawamoto:20:SoSyM} is the first attempt to construct a modal logic to describe statistical properties of hypothesis testing.
They introduce a belief modality weaker than S5 and interpret it in a Kripke model where an accessibility relation is defined using a statistical distance between possible worlds.
However, these logics cannot 
reason about the procedures of statistical methods.

\noindent
\textit{Belief Hoare Logic.}
Belief Hoare logic (BHL)~\cite{Kawamoto:21:KR,Kawamoto:24:AIJ} is a program logic with an epistemic modality for statistical belief.
Using this logic, we can derive the correctness of a hypothesis testing program (\Sec{sub:BHL}).
Our verification tool, \StatWhy{}, uses BHL as its theoretical foundation to produce a proof tree for the correctness of hypothesis testing programs within the tool.
To develop \StatWhy{},
we implemented several constructs specific to BHL using the WhyML language---the intermediate language used within \WhyT{} framework.
This allows verification conditions generated by \StatWhy{} to be discharged by off-the-shelf SMT solvers.

\noindent
\textit{Program Verification Tools.}
Various tools used for specifying the preconditions and postconditions of each function and statically verifying their correctness;
for example, Frama-C~\cite{DBLP:conf/sefm/CuoqKKPSY12} for C programs;
Dafny~\cite{DBLP:conf/lpar/Leino10} for imperative programs that compile to Boogie~\cite{leino2008boogie}; 
ESC/Java~\cite{DBLP:journals/sigplan/FlanaganLLNSS13} 
and KeY~\cite{KeYBook2016} for Java programs.
To the best of our knowledge, no attempt has been made to apply these tools for verifying the correct usage of statistical methods in programs.

\noindent
\textit{The Verification Frameworks Used in Our Tool.}
\StatWhy{} is implemented as an extension of \Cameleer{}~\cite{Pereira:20:CAV}, a verifier for OCaml programs built on top of the \WhyT{} framework~\cite{Filliatre:13:ESOP}.
\Cameleer{} works as a translator from OCaml to the simple functional programming language WhyML.
The translated WhyML code is then verified by \WhyT{}.
\StatWhy{} extends \Cameleer{} to verify the correctness of hypothesis testing programs written in OCaml by incorporating the constructs and inference rules of BHL.

\noindent
\textit{Verification of Statistical Algorithms.}
From a broader perspective, a number of studies have investigated the numerical accuracy of statistical algorithms~\cite{Lesage:85:CSDA}, the formal verification of randomized algorithms~\cite{Kozen:85:JCSS,Morgan:96:TPLS}, and the PAC verification~\cite{Goldwasser:21:ITCS,Mutreja:23a:COLT} for approximately checking the correctness of statistical algorithms.
Furthermore, formal methods have been used to verify a certain guarantee of the correctness of statistical machine learning models~\cite{Seshia:22:CACM}.
However, no prior work has provided a formal method tool for specifying or verifying the correct usage of statistical methods rather than the correctness of statistical algorithms.

\noindent
\textbf{Plan of the paper.}
\label{sub:plan}
The rest of the paper is organized as follows.
In \Sec{sec:background}, we review basic notions in hypothesis testing and belief Hoare logic (BHL).
In \Sects{sec:design} and~\ref{sec:implementation}, we present \StatWhy{}'s design and implementation, respectively.
In \Sec{sec:applications}, we show examples of \StatWhy{} being applied to common errors in hypothesis testing.
In \Sec{sec:conclude}, we present our final remarks.

\section{Background}
\label{sec:background}

\label{sub:HT}

\emph{Statistical hypothesis testing}~\cite{Arbuthnot:1710,Hogg:2013:book-intro} is a method for inferring information about an unknown population $x$ from a dataset $y$ that has been sampled from the population~$x$.
In a hypothesis test, the \emph{null hypothesis} $\phi_0$ is a claim that we wish to test, while the \emph{alternative hypothesis} $\phi_1$ is a claim that we will accept if the null hypothesis is rejected.
The goal of a hypothesis test is to determine whether we have sufficient evidence to reject the null hypothesis $\phi_0$ in favor of the alternative hypothesis $\phi_1$.

\begin{example}[$t$-test for a population mean] \label{eg:t-test}
For a population $x$ following a normal distribution with an unknown mean,
the \emph{$t$-test for the population mean} is a hypothesis test to check whether the unknown mean $\mean(x)$ differs from a certain value $\mu_0$ specified in the null hypothesis.

In the $t$-test, we want to show the alternative hypothesis $\phi_1 \eqdef (\mean(x) \neq \mu_0)$ by rejecting the null hypothesis $\phi_0 \eqdef (\mean(x) = \mu_0)$.
First, we calculate the \emph{$t$-test statistic} from a dataset $y$:
$t(y) \detassign \frac{\mean(y) - \mu_0}{\nicefrac{s}{\sqrt{n}}}$
where $n$ is the sample size of $y$ and $s$ is a sampled standard deviation, i.e.,
$s \eqdef \sqrt{\frac{\sum_{i=1}^{n} (y_i - \mean(y))^2}{n-1}}$.
This statistic is compared to Student's $t$-distribution with $n-1$ degrees of freedom
(i.e., the distribution of the $t$-statistic $t(y)$ when $y$ is normally distributed).
Specifically, we calculate the \emph{$p$-value}:
$\Pr_{d \sim \Normal(\mu_{\ppl},\sigma^2)^n}[\, |t(d)| > |t(y)| \,]$
under the  null hypothesis $\phi_0$.
For a smaller $p$-value, the dataset $y$ is far from what we expect under the null hypothesis that $\mean(x) = \mu_0$.
Hence, in the $t$-test, if the $p$-value is smaller than a certain threshold (e.g., $0.05$), we reject the null hypothesis $\phi_0$ and accept the alternative hypothesis $\phi_1$,
i.e., $\mean(x) \neq \mu_0$.

We remark that this $t$-test requires that the population $x$ should follow a normal distribution.
If this requirement is not satisfied, the use of the $t$-test is inappropriate and its result may be incorrect.
Therefore, analysts need to check this requirement in some way.
Since they cannot mathematically prove this requirement on the unknown population $x$, 
they use their background knowledge about the population $x$ and check approximately whether the dataset $y$ at hand (rather than the unknown population $x$ itself) follows a normal distribution.
\end{example}

\label{sub:BHL}

\emph{Belief Hoare logic} (\emph{BHL})~\cite{Kawamoto:21:KR,Kawamoto:24:AIJ} is a program logic equipped with epistemic modal operators for the statistical beliefs acquired via hypothesis testing.
The epistemic modal logic used in BHL is defined by:
\begin{align*}
\phi \,\mathbin{::=}\,\,&
\eta(u_1, \ldots, u_n) \mid
\neg \phi \mid \phi \land \phi \mid \Know\phi 
\mid \KnowXx{\le \epsilon}{y, A} \phi
\end{align*}
for a predicate symbol $\eta$, terms $u_1, \ldots, u_n$, a dataset $y$, a hypothesis test $A$, 
and a $p$-value $\epsilon$.
The knowledge modality $\Know$ is defined in the S5 modal logic system with axioms T, 4, and 5.
Intuitively, $\Know \phi$ represents that we know $\phi$.
The epistemic possibility $\Possible$ is defined as usual by 
$\Possible\phi \eqdef \neg \Know \neg \phi$.
$\KnowXx{\le \epsilon}{y, A} \phi$ represents that by a hypothesis test $\alg$ on a dataset $y$, we believe $\phi$ with a $p$-value $\alpha \le \epsilon$.

The semantics of this epistemic logic is based on a Kripke model~\cite{Kawamoto:21:KR,Kawamoto:24:AIJ}.
The satisfaction of an epistemic formula $\phi$ in a possible world $w$ is denoted by $w \models \phi$ and is defined straightforwardly in a Kripke model 
where each possible world is equipped with a test history that is updated by performing hypothesis tests.

In the framework of BHL, we express a procedure for hypothesis testing as a program $C$ using a programming language.
Then, we use epistemic modal logic to
describe the requirements for the hypothesis tests as a precondition formula,
e.g.,
$\psiPre \eqdef \sampled{y}{\Normal(\mu_0, \sigma^2)}{}\, \land \Possible \phi_{1} \land \kappaX{\emptyset}$,
where 
the atomic formula
$\sampled{y}{\Normal(\mu, \sigma^2)}{}$
represents that a dataset $y$ is sampled from the population that follows a normal distribution $\Normal(\mu, \sigma^2)$ with an unknown mean $\mu$ and an unknown variance $\sigma^2$.
The modal formula $\Possible \phi$ represents that, before conducting the hypothesis test, we have the \emph{prior belief} that the alternative hypothesis $\phi$ \emph{may be} true.
The formula $\kappaX{\emptyset}$ represents that no hypothesis test has been conducted previously.

The statistical belief we acquire from the hypothesis test is specified as a \emph{postcondition} formula,
e.g., $\phiPost \eqdef \KnowXx{\le0.05}{y, A}\! \phi_{1}$,
representing that by a hypothesis test $\alg$ on the dataset $y$,
we believe $\phi$ with a $p$-value $\alpha \le 0.05$.
Since the result of the hypothesis test may be wrong, we use the statistical belief modality $\KnowXx{\le0.05}{y, A}$ instead of the knowledge modality $\Know$.

Finally, we combine all the above and describe the whole statistical inference as a \emph{judgment}
$\env\vdash\triple{ \psiPre }{ C }{ \phiPost }$,
representing that whenever the precondition $\psiPre$ is satisfied, the execution of the program $C$ results in the satisfaction of the postcondition $\phiPost$.
By deriving this judgment using derivation rules in BHL, we conclude that the program $C$ for the statistical inference results in the statistical belief $\phiPost$ whenever the requirement $\psiPre$ is satisfied.

BHL has been applied only to pen-and-paper analyses of a few simple examples of statistical hypothesis testing in previous work~\cite{Kawamoto:21:KR,Kawamoto:24:AIJ} and has not yet been used to build a computer-aided verification tool.

\section{The Design of \StatWhy{}}
\label{sec:design}

\subsection{Running Examples}
\label{sec:illustrating}

\begin{wrapfigure}[4]{r}{0.59\textwidth}
  \vspace{-5.0em}
\begin{lstlisting}[frame = single, label=code1ocaml]
 let p = exec_ttest_1samp t_n 1.0 d Two
 (*@  requires  sampled d t_n
      ensures  (World (!st) interp) |= StatB p fmlA  *)
\end{lstlisting}
  \vspace{-1.0em}
\caption{An OCaml program that calls a $t$-test command for a mean of a population. \label{fig:illustrating:OCaml}}
\end{wrapfigure}

\noindent
\textbf{Simple Example.}
We present the main idea of our formal specification and automated verification method
using the program in \Fig{fig:illustrating:OCaml}.
This program shows an OCaml source code that executes a command ${\tt exec\_ttest\_1samp}$ for the one-sample $t$-test (\Eg{eg:t-test} in \Sec{sub:HT}) with an alternative hypothesis \texttt{fmlA} 
(e.g., representing $\texttt{mean(t\_n) != 1.0}$).

To specify the requirements and the interpretation of this $t$-test command, a programmer writes the precondition in the  \texttt{requires} clause and the postcondition in the \texttt{ensures} clause using the specification language \Gospel{}~\cite{Chargueraud:19:FM}.

In this code, 
\texttt{sampled} is a predicate defined in WhyML, and
the precondition 
\texttt{sampled d t\_n}
expresses that the dataset \texttt{d} has been sampled from a population~that has a normal distribution type \texttt{t\_n} with an unknown mean and an unknown variance.
In the WhyML implementation of BHL, a dataset is implemented as a record with a field storing the distribution type of the population.

The postcondition 
\texttt{(World (!st) interp) |= StatB p fmlA}
represents the interpretation of the result of the hypothesis test.
Specifically, we obtain a \emph{statistical belief}---denoted by the logical formula \texttt{StatB p fmlA}---that an alternative hypothesis \texttt{fmlA} holds with a $p$-value \texttt{p}, in the possible world 
\texttt{(World (!st) interp)}
equipped with the record \texttt{st} of all hypothesis tests executed so far.
This postcondition encodes the satisfaction of the epistemic formula $w \models \KnowXx{\le \texttt{p}}{y, A} \texttt{fmlA}$ in the world $w$ in the Kripke model addressed in \Sec{sec:background}.

By applying \StatWhy{} to this source code, 
the program verification fails because other requirements are missing in the precondition.
However, 
since \WhyT{} finds the failure to discharge the verification conditions corresponding to such requirements,
the tool user can easily find out the missing requirements.

We remark that the specification of \texttt{exec\_ttest\_1samp} is defined in \StatWhy{} using WhyML so that it
(1) checks the requirements for the hypothesis test in the precondition,
(2) asserts the conclusion implied by the hypothesis test in the postcondition, and 
(3) updates the test history \texttt{st} with the test result 
(\Fig{fig:spec-one-sample-t}).

\begin{figure}[t]
    \centering
    \begin{lstlisting}[frame = single]
 val exec_ttest_1samp (p:population) (mu:real) (y:dataset (list real)) (alt:alternative):real
  ...
  requires {
   match p with
   | (NormalD _ _) -> sampled y p /\ ... /\
     match alt with
     | Two -> (World !st interp) |= Possible ((mean p) <' (const_term mu)) /\
              (World !st interp) |= Possible ((mean p) >' (const_term mu)) ...
  }
  ensures {
   pvalue result /\
   let h = match alt with
     | Two -> (mean p) $!= (const_term mu) ...
   end in !st = Cons ("ttest_1samp", h, Eq result) !(old st)
  }
    \end{lstlisting}
    \vspace{-2ex}
    \caption{The specification of \texttt{exec\_ttest\_1samp}}
    \label{fig:spec-one-sample-t}
\end{figure}

In the \texttt{requires}-clause, the two-tailed (\texttt{TWO}) $t$-test requires the prior belief that both tails $\mean(p) < \mu$ and $\mean(p) > \mu$ are possible.
In the \texttt{ensures}-clause, the test result consisting of the test name, the hypothesis $\texttt{h}$, and the $p$-value $\texttt{result}$ is added to the test history $\texttt{st}$.
This specification of the two-tailed $t$-test encodes an instance of the following inference rule in BHL~\cite{Kawamoto:21:KR,Kawamoto:24:AIJ}:
\begin{gather*}
\dfrac{\begin{array}{c}
  \Gamma \models \psi \implies (\cmpds{} \land \Possible \phiL \land \Possible \phiU)
\end{array}
}{
  \env \vdash
  \triple{
  \psi \land \kappaX{\emptyset}
  }{\alpha := f_{\!\alg}(y)}{
  \psi \land \kappaX{y,\alg} \land \KnowXx{\alpha}{y,\alg} (\phiL \lor \phiU)
  }
}
\tag{$\axTT{}$}
\end{gather*}%
where the precondition $\psi$ includes the assumption $\cmpds{}$ on the population distribution and the prior beliefs on the two tails $ \Possible \phiL$ and $\Possible \phiU$;
the postcondition updates the empty test history $\kappaX{\emptyset}$ to the history $\kappaX{y,\alg}$ recording the test result.

\noindent
\textbf{$P$-Value Hacking.}
\Fig{fig:p-value:hacking} is another example presenting how \StatWhy{} detects an error in the code conducting
the \emph{p-value hacking} (a.k.a. \emph{data dredging}), a technique to manipulate statistical analyses to obtain a lower $p$-value.

\begin{figure}[t]
\centering
\begin{lstlisting}[frame = single]
 let ex_hack trial1 trial2 =
   let p1 = exec_ttest_1samp ppl_new 1.0 trial1 Two in
   let p2 = exec_ttest_1samp ppl_new 1.0 trial2 Two in
   let p = min p1 p2 in (* This is INCORRECT *)
   (p1, p2, p)
 (*@ (p1, p2, p) = ex_hack trial1 trial2
   requires
     is_empty (!st) /\ sampled trial1 ppl_new /\ sampled trial2 ppl_new /\
     (World (!st) interp) |= Possible h_new_l /\ (World (!st) interp) |= Possible h_new_u
   ensures
     (Leq p = compose_pvs h_new !st
       && (World !st interp) |= StatB (Leq p) h_new) *)
\end{lstlisting}
\vspace{-2ex}
\caption{An OCaml program that performs the $p$-value hacking. \label{fig:p-value:hacking}}
\end{figure}

In this program, we execute a $t$-test \texttt{exec\_ttest\_1samp} on a dataset $\texttt{trial1}$ and another on another dataset $\texttt{trial2}$.
Given the $p$-values $\texttt{p1}$ and $\texttt{p2}$ for these two $t$-tests, we should report $\texttt{p1} + \texttt{p2}$ as the $p$-value of these experiments.
However, this program 
reports only the experiment showing the lower $p$-value (i.e., $\texttt{min p1 p2}$) by ignoring the other showing higher $p$-value.
By ensuring that all hypothesis tests are described in the program, \StatWhy{} can automatically check whether the $p$-values are correctly calculated, thus preventing $p$-value hacking.

We remark that
in the precondition, the atomic expression \texttt{is\_empty (!st)} 
with the dereference operator `\texttt{!}'
represents that the test history \texttt{st} is empty;
i.e., no hypothesis test has performed before.
The expression $\texttt{sampled trial1 ppl\_new}$ represents that the dataset $\texttt{trial1}$ is sampled from a population $\texttt{ppl\_new}$.
For specific predicates such as $\texttt{is\_empty}$ and $\texttt{sampled}$, we can use abbreviated expressions where ``$\texttt{(World !st interp) |= }$'' is omitted for the sake of simplicity.
In the postcondition, $\texttt{compose\_pvs h\_new !st}$ obtains the correct $p$-value from the test history $\texttt{st}$, which is found to be different from the $p$-value $\texttt{p}$ incorrectly calculated in this program.

\subsection{More Details on Specifications}

The latest version of \StatWhy{} can automatically verify the correctness of specifications written in the \WhyML{} language~\cite{Filliatre:13:ESOP}.
It can also verify programs written in the subset of OCaml supported by \Cameleer{}~\cite{Pereira:20:CAV}, a verifier for OCaml programs.  Specifically, \Cameleer{} covers the core OCaml language except for several features, including object-oriented programming, generalized algebraic data types (GADTs), and polymorphic variants.

\StatWhy{} requires minimal modifications to the source code.
Programmers need to insert annotations into the OCaml program to describe the requirements and interpretations for hypothesis testing.
More specifically, a requirement (respectively, interpretation) for a hypothesis testing command is expressed as a logical formula written in the \Gospel{} language~\cite{Chargueraud:19:FM}, representing a precondition (respectively, postcondition) for the command.

To describe these annotations
in \Gospel{}, we introduce types for \emph{terms}, \emph{atomic formulas}, and \emph{logical formulas} of belief Hoare logic (BHL) as follows.
\begin{lstlisting}
  type term = RealT real_term | PopulationT population | ...
  type atomic_formula = Pred psymb (list term)
  type formula = Atom atomic_formula | Not formula 
               | Conj formula formula | Disj formula formula
               | Possible formula | Know formula | StatB pvalue formula | ...
\end{lstlisting}
where a term can express a real number and a population; an atomic formula consists of a predicate symbol (e.g., \texttt{is\_normal}) and a list of terms; 
a formula is built using \texttt{Possible}, \texttt{Know}, and \texttt{StatB},
each corresponding to the modal epistemic operators $\Possible$, $\Know$, and $\KnowXx{\le \epsilon}{y, A}$, respectively, shown in \Sec{sec:background}.

In the WhyML grammar, an atomic expression is of the form 
$
\texttt{World (!st)} \allowbreak \texttt{interp } \texttt{|= formula}
$
representing that the formula $\texttt{formula}$ is satisfied in the possible world equipped with the test history $\texttt{st}$ (i.e., the record of all hypothesis tests executed so far) and the interpretation $\texttt{interp}$ of private-variables in the Kripke model for BHL~\cite{Kawamoto:21:KR,Kawamoto:24:AIJ}. 
For the non-modal formulas using only specific predicates (e.g., $\texttt{is\_empty}$ or $\texttt{sampled}$), we allow an abbreviation that can omit ``$\texttt{World (!st) interp |=}$'' from an expression.
We can also use function symbols (e.g., $\mean$ and \texttt{ppl}) to simplify expressions.

For clarity in hypothesis testing specifications, programmers can use abbreviation operators.
Since hypothesis testing programs often involve repeated comparisons among multiple groups of data, 
\StatWhy{} provides a set of folding operations to simplify the repetition of similar conditions in specifications.
In particular, folding operators can be used to briefly describe the iteration of the hypothesis tests that compare all combinations of groups in multiple comparison.

\begin{figure}[t]
\begin{lstlisting}[frame = single, label=code1whyml]
 use ocamlstdlib.Stdlib
 let function p =
   [@vc:white_box]
   (begin
      requires { sampled d t_n }
      returns { p -> (World (!st) interp) |= (StatB p fmlA) }
      exec_ttest_1samp t_n 1.0 d Two 
    end)
\end{lstlisting}
\vspace{-2ex}
\caption{A WhyML program calling a $t$-test command for a mean of a population. \label{fig:transform}}
\end{figure}

\subsection{Verification of Statistical Programs}
\label{subsec:verification}

To verify a given OCaml program,
\StatWhy{} first transforms it into a program written in the WhyML language~\cite{Filliatre:13:ESOP}.
This preprocessing is performed using our extension of \Cameleer{}~\cite{Pereira:20:CAV}, a static verifier for OCaml.
For example, the OCaml program in \Fig{fig:illustrating:OCaml} is transformed into the 
\WhyML{} program in \Fig{fig:transform}.

Next, \StatWhy{} proves the correctness of a WhyML program using the \WhyT{} platform~\cite{Filliatre:13:ESOP}.
Specifically, the tool internally synthesizes a proof tree using the proof rules of Belief Hoare logic and derives the verification condition for the program.
Then, it automatically discharges these conditions by using external SMT solvers, e.g., \cvcFive{}~\cite{DBLP:conf/tacas/BarbosaBBKLMMMN22} or Z3~\cite{DBLP:conf/tacas/MouraB08}.
If the verification succeeds, \StatWhy{} outputs a proof tree that attests to the correctness of the program.
If the verification fails or times out, the tool reports the failure.
Even in that case, the tool users can identify any missing or incorrect requirements and interpretations for statistical methods so that they can re-specify the requirements and interpretations.\arxiv{; see Appendix~\ref{sub:Why3} for more detail}.

The verification process using \StatWhy{} guarantees the following correctness.
If \StatWhy{} successfully verifies a program, for any function $\texttt{f}$ defined as $\texttt{let f d1 ... dn = e}$, annotated with a precondition $\psiPre$ and a postcondition $\phiPost$, the judgment
$\triple{\psiPre}{[v_1/\texttt{d1}, \ldots , v_n/\texttt{dn}]\, \texttt{e}}{\phiPost}$ holds for any value $v_i$ of type $\texttt{d}i$, assuming the soundness of the \WhyT{} framework.
By the soundness of BHL, if the expression $\texttt{e}$ is evaluated under the environment satisfying the precondition $\psiPre$, then the resulting environment satisfies the postcondition $\phiPost$.

We remark that \StatWhy{} focuses on automatically verifying the \emph{procedure} and the \emph{annotations} in a statistical program 
by assuming the correctness of the implementation of each hypothesis testing method used in the procedure as a subroutine.
In other words, our automated verification tool 
only checks that a program correctly uses hypothesis testing functions.
Technically, \StatWhy{} checks whether the preconditions and the postconditions of hypothesis testing functions are satisfied when the program calls these functions as subroutines.
By using \StatWhy{}, programmers are encouraged to pay attention to the requirements, the interpretations, and the choices of hypothesis testing methods.

We also remark that \StatWhy{} verifies a statistical program only under the assumption that all requirements about an unknown true population are satisfied
(e.g., a population follows a normal distribution).
Thus, such an assumption needs to be checked externally; for instance,
the analysts are responsible for judging whether the population appears to approximately follow the normal distribution\footnote{There are hypothesis testing methods for checking the normality approximately. 
Such tests are applied to the \emph{dataset} (instead of the actual \emph{population}) and cannot prove the normality of the population mathematically.}, possibly using background knowledge about the population.

\section{The Implementation of \StatWhy{}}
\label{sec:implementation}

 In this section, we explain and discuss the implementation of the \StatWhy{} tool.
 More details on the tool is available in the user documentation~\cite{Kawamoto:24:UD}.

\subsection{The Architecture of \StatWhy{}}
\label{subsec:components}

\Fig{fig:statwhy} shows the architecture of \StatWhy{}.
To specify the requirements and interpretations of hypothesis tests as preconditions and postconditions, \StatWhy{} has modules for real number arithmetic, basic statistics, and individual hypothesis testing commands (e.g., $t$-test)
that cover most of the popular hypothesis testing methods~\cite{Kanji:06:book:100stat}.
To reason about the interpretation of hypothesis testing, the tool also has modules for BHL~\cite{Kawamoto:21:KR,Kawamoto:24:AIJ}, an epistemic logic with statistical belief explained in \Sec{sub:BHL}, 
and for the record of the hypothesis tests performed so far.

\begin{wrapfigure}[10]{r}{0.55\textwidth}
  \vspace{-2.0em}
\scalebox{0.55}{
  \begin{tikzpicture}[
    font=\footnotesize,
    node distance=1em,
    box/.style={
      draw,
      fill=blue!10,
      rounded corners,
      align=center,
    },
    bigbox/.style={
      draw,
      thick,
      rounded corners,
      inner sep=0.4em
    }
  ]

\node[box, align=left] (pre) {\footnotesize \textbf{Pre-cond.}:\\ Req. for\\ stat.};
\node[box, align=center, below=0.2em of pre] (main) {\footnotesize \textbf{Main}\\ \textbf{program}};
\node[box, align=left, below=0.2em of main] (post) {\footnotesize \textbf{Post-cond.}:\\ Interp. of\\ stat.};
\node[
  bigbox,
  label={[font=\small]above:\textbf{OCaml code}},
  fit=(pre)(main)(post)
] (ocaml) {};
\node[
  bigbox,
  label={[font=\small]above:\textbf{StatWhy}},
  minimum width=6.3cm,
  minimum height=4.6cm,
  right=2.4em of ocaml,
] (statwhy) {};
\node[box] (cameleer) at ([xshift=3em,yshift=-0.8em] statwhy.west) {\textbf{Cameleer}};
\node[box, above=0.2em of cameleer, xshift=0em] (msl) {\scriptsize Mod. for\\ \scriptsize stat. \& logic};
\node[box, text width=2.5cm, right=4em of cameleer] (why3) {Why3 Platform};
\node[box, text width=2.5cm, above=0.1em of why3, xshift=0.0cm] (mepi) {\scriptsize Mod. for epistemic logic};
\node[box, text width=2.6cm, above=0.1em of mepi, xshift=0.0cm] (mhyp) {\scriptsize Mod. for hypothesis test commands (Z-test, T-test, \dots)};
\node[box, text width=2.5cm, below=2em of why3] (smt) {\footnotesize SMT solvers (Z3, CVC5, \dots)};
\draw[->, thick] (cameleer.east) -- node[above, text width=4em, align=center]{\scriptsize Trans. to WhyML} (why3.west);
\draw[->, thick] (why3.south) -- node[right, text width=2em]{\scriptsize Discharge VCs} (smt.north);
\draw[->, thick] (smt.north) -- (why3.south);
\draw[->, thick] (ocaml.east) -- node[above, midway]{\scriptsize input} (statwhy.west);
\node[box, right=3.8em of why3] (proof) {Proof};
\draw[->, thick] (why3.east) -- node[above]{\scriptsize output} (proof.west);
\end{tikzpicture}
}
\caption{The architecture of the \StatWhy{} tool.}
\label{fig:statwhy}
\end{wrapfigure}

Since the goal of our verification tool is to ensure the correct usage of the hypothesis testing methods in programs, the \StatWhy{} tool reasons about $p$-values in hypothesis testing, which requires basic reasoning about probabilities. 
Specifically, we extended \Cameleer{} so that it can access the real number arithmetic formalized in \WhyT{} standard library,
and added basic statistics functions (e.g., mean) and their lemmas.
To reason about $p$-values appearing in epistemic formulas in simplifying verification conditions, we added \WhyT{} lemmas about the composition of $p$-values under multiple-tests and about the comparisons between $p$-values.
With these lemmas in the \StatWhy tool{}, we avoid the need for SMT solvers to handle probability computations, while maintaining the soundness and correctness in the statistical context.

To accept OCaml programs as input, \StatWhy{} internally calls our extension of \Cameleer{}~\cite{Pereira:20:CAV} to translate an OCaml program to a WhyML program. 
Then the tool verifies a WhyML program by using the \WhyT{} platform.

\subsection{Extension of \Cameleer{}}
\label{sub:extend-cameleer}

The programs given to \StatWhy{} share several peculiar features.
For example, \StatWhy{} specifications often involve repeated comparisons among multiple groups of data, which are expressed using folding operations.
Furthermore, the programs often involve the array structure of records of hypothesis tests executed so far.
We have found that SMT solvers often get stuck if we try to discharge the verification conditions that involve such folding operations.

To improve the performance of \StatWhy{}, we implemented a custom proof strategy---a combination of proof tactics and transformations---that exploits these characteristics of hypothesis testing programs to accelerate the proof search.
Our strategy first applies \WhyT{}'s default proof strategies (e.g., \texttt{split\_vc} for splitting conjunctive verification conditions into smaller ones and \texttt{compute\_in\_goal} for applying computations and simplifications to proof goals).
These invocations of the proof strategies are interleaved with calls to SMT solvers, whose timeouts are set to small values.
If these applications of the default proof strategies fail to discharge the VCs, then we apply aggressive transformation strategies that unfold the definitions of the functions and predicates defined in \StatWhy{}.

These extensions are implemented as follows.
We added WhyML modules for \StatWhy{} to the standard library of \WhyT{} at the installation of our extension of \Cameleer{}.
We also extended \Cameleer{} so that \texttt{Uterm}, the module for OCaml untyped terms, and \texttt{Why3ocaml\_driver}, the module for translation from OCaml to WhyML, can handle floating-point numbers.
We also extended the \WhyT{} plugins provided by \Cameleer{} by adding \texttt{plugin/cameleerBHL.ml} and modifying \texttt{plugin/plugin\_cameleer.ml} to handle the algebraic data types in \StatWhy{}.

\section{Case Studies on Common Errors in Hypothesis Testing}
\label{sec:applications}
\label{sub:multi-comparison}
We present examples to demonstrate how \StatWhy{} can be used to avoid common errors in a variety of popular hypothesis testing programs.

\begin{figure}[t]
\centering
\begin{lstlisting}[frame = single]
 let cmp_with_existing_drugs d_new d_drug1 d_drug2 =
   let p_drug1 = exec_ttest_ind_eq ppl_new ppl_drug1 (d_new, d_drug1) Up in
   let p_drug2 = exec_ttest_ind_eq ppl_new ppl_drug2 (d_new, d_drug2) Up in
   p_drug1 +. p_drug2
  (*@ p = testing d_new d_drug1 d_drug2
     requires
       is_empty (!st) /\ non_paired d_new d_drug1 /\ non_paired d_new d_drug2 /\
       sampled d_new ppl_new /\ sampled d_drug1 ppl_drug1 /\ sampled d_drug2 ppl_drug2 /\
       (World (!st) interp) |= Possible h_new_drug1 /\
       (World (!st) interp) |= Not (Possible h_new_drug1_c) /\
       (World (!st) interp) |= Possible h_new_drug2 /\
       (World (!st) interp) |= Not (Possible h_new_drug2_c)
     ensures
       (Leq p) = compose_pvs (Disj h_new_drug1 h_new_drug2) !st &&
       (World !st interp) |= StatB (Leq p) (Disj h_new_drug1 h_new_drug2) *)
\end{lstlisting}
\vspace{-2ex}
\caption{An OCaml program that performs multiple comparison. \label{fig:multiple-comparison}}

\end{figure}

\noindent
\textbf{Multiple Comparison Problems.}
Analysts occasionally make mistakes in computing the $p$-value in comparing more than two groups,
which is called a \emph{multiple comparison problem}~\cite{Bretz:10:book}.

Let us consider an experiment comparing the efficacy of a new drug with that of two existing drugs \drugA{} and \drugB{}.
We conduct two one-tailed $t$-tests: one comparing the new drug with \drugA{} and another with \drugB{}.
Then the alternative hypotheses \texttt{h\_new\_drug1} and \texttt{h\_new\_drug2} for these tests represent that the new drug has a better efficacy than \drugA{} and \drugB{}, respectively.
For the $p$-values $p_1$ and $p_2$ of these two tests, 
the $p$-value $p$ for the combined test with the disjunctive hypothesis
$\texttt{h\_new\_drug1} \lor \texttt{h\_new\_drug2}$
satisfies $p \le p_1 + p_2$, which is known as \emph{Bonferroni correction}.
In contrast, the $p$-value $p$ for the conjunctive hypothesis 
$\texttt{h\_new\_drug1} \land \texttt{h\_new\_drug2}$
satisfies $p \le \min(p_1, p_2)$.

\StatWhy{} can automatically check that the program 
in \Fig{fig:multiple-comparison}
correctly calculates the $p$-value of the disjunctive hypothesis \texttt{Disj h\_new\_drug1 h\_new\_drug2}.

\begin{table}[t]
\centering
\caption[number-of-hypotheses]{
The execution times (sec) for verifying hypothesis testing programs with practical numbers of disjunctive (OR) and conjunctive (AND) hypotheses.
\label{tab:num-hyp}
}
~\\[0.0ex]
{
\tabcolsep=4pt
\renewcommand{\arraystretch}{1}
\begin{small}
\begin{tabular}{ccccccccccc} \toprule
\#hypotheses & 2 & 3 & 4 & 5 & 6 & 7 & 8 & 9 & 10 \\ \midrule
OR  & 8.77 & 8.89 & 8.84 & 8.94 & 9.01 & 9.01 & 9.04 & 9.16 & 9.23 \\
AND & 8.82 & 8.72 & 8.86 & 8.98 & 8.95 & 9.03 & 9.11 & 9.17 & 9.46 \\
\bottomrule
\end{tabular}
\end{small}
}
\end{table}

\begin{table}[t]
    \vspace{-1.0em}
    \centering
    \caption[multiple-comparison-methods]{
    The execution times (sec) for various multiple comparison methods.
    \#groups (respectively, \#comparisons) represents the (practical) number of groups (respectively, combinations of groups) compared in the testing.
    }
    \label{tab:multiple-comparison-methods}
    {
    \tabcolsep=4pt
    \renewcommand{\arraystretch}{1}
    \begin{small}
    \begin{tabular}{llrrrrrr}
    \toprule
    & & \multicolumn{6}{c}{\#groups} \\
    \cmidrule(lr){3-8}
    \multicolumn{1}{c}{Test methods} & \multicolumn{1}{c}{Metric} & \multicolumn{1}{c}{2} & \multicolumn{1}{c}{3} & \multicolumn{1}{c}{4} & \multicolumn{1}{c}{5} & \multicolumn{1}{c}{6} & \multicolumn{1}{c}{7} \\
    \midrule
    Tukey's HSD test & Times (sec)      & 0.37 & 9.09 & 9.33 & 9.81 & 15.27 & 16.39 \\
                     & \#comparisons & 1    & 3    & 6    & 10   & 15    & 21 \\
    Dunnett's test   & Times (sec)      & 0.48 & 8.98 & 9.17 & 9.61 & 9.62  & 9.77 \\
                     & \#comparisons & 1    & 2    & 3    & 4    & 5     & 6 \\
    Williams' test   & Times (sec)      & 0.48 & 8.90 & 9.04 & 9.16 & 9.23  & 9.58 \\
                     & \#comparisons & 1    & 2    & 3    & 4    & 5     & 6 \\
    Steel-Dwass' test & Times (sec)     & 0.44 & 9.05 & 9.43 & 9.76 & 15.10 & 16.24 \\
                     & \#comparisons & 1    & 3    & 6    & 10   & 15    & 21 \\
    Steel's test     & Times (sec)      & 0.49 & 8.79 & 8.92 & 9.11 & 9.43  & 9.74 \\
                     & \#comparisons & 1    & 2    & 3    & 4    & 5     & 6 \\
    \bottomrule
    \end{tabular}
    \end{small}
    }
\end{table}

\noindent
\textbf{Scalability of \StatWhy{}}
We evaluated the scalability of the performance of the program verification using \StatWhy{} to (i) the complex hypotheses and (ii) the larger number of compared groups.
For the evaluation, we conducted experiments
on a MacBook Pro with Apple M2 Max CPU and 96 GB memory using the external SMT solver \cvcFive{} 1.0.6.

\Tbl{tab:num-hyp} shows the execution times for \StatWhy{} to verify hypothesis testing programs for practical numbers of disjunctive/conjunctive hypotheses.
These experiments took roughly the same amount of time for a larger number of hypotheses.
\Tbl{tab:multiple-comparison-methods} provides the execution times for 
the most common multiple comparison methods described in standard textbooks.
The numbers of groups compared in the experiments are practical but challenging, as the number of comparisons grows rapidly with the number of groups.
The verification of these programs is efficient, since our proof strategy (\Sec{sec:implementation}) accelerates the proof search for programs with folding operations and test histories.

\section{Conclusion}
\label{sec:conclude}
We proposed a tool-assisted method for formally specifying and automatically verifying the correctness of statistical programs. 
In particular, we presented the \StatWhy{} tool for automatically checking whether the programmers have properly specified the requirements and the interpretations of the statistical methods.

In future work, 
we will extend \StatWhy{} to verify the procedures for 
power analyses and interval estimation
and to deal with
other types of statistical methods and
other programming languages, e.g., a subset of Python.
We also plan to work on the formal verification of the correctness of the implementation of each hypothesis testing function 
called from statistical software as a subroutine.

\ifanonymous
\else
\begin{credits}
\subsubsection{\ackname} 
The authors are supported by JSPS KAKENHI Grant Number JP24K02924, Japan.
Yusuke Kawamoto is supported by JST, PRESTO Grant Number JPMJPR2022, Japan.
Kohei Suenaga is supported by JST CREST Grant Number JPMJCR2012, Japan.
\subsubsection{\discintname}
The authors have no competing interests to declare that are
relevant to the content of this article. 
\end{credits}
\fi
\bibliographystyle{splncs04}
\bibliography{short,short-stat,sv}

\arxiv{%
\appendix
\input{_appendix}
}%

\end{document}